\documentclass[%
 reprint,
 superscriptaddress,
 amsmath,amssymb,
 aps,
 pre,
 showkeys, % Maybe not?
 floatfix,
]{revtex4-2}

\usepackage{graphicx}       % Include figure files
\usepackage[caption=false]{subfig}
\usepackage{dcolumn}        % Align table columns on decimal point
\usepackage{bm}             % Bold math
\usepackage{hyperref}       % Add hypertext capabilities
\usepackage{xcolor}         % Colored text
\usepackage{physics}        % Physics notation
\usepackage{multirow}

\iftrue
    \newcommand{\todo}[1]{{\color{red}[#1]}} % Marking todo
\else
    \newcommand{\todo}[1]{}               % Default definition
\fi

\renewcommand{\vec}[1]{\boldsymbol{#1}}

\DeclareMathSymbol{\shortminus}{\mathbin}{AMSa}{"39}

\newcommand{\boltz}{k_{\text{B}}}
\newcommand{\bohr}{a_{\text{B}}}
\renewcommand{\order}[1]{\mathcal{O}\left(#1\right)}

\begin{document}

\preprint{APS/123-QED}

%%%%%%%%%%%%%%%%%%%isFRONTMATTER %%%%%%%%%%%%%%%%%%%%%

\title{Modelling of warm dense hydrogen via explicit real time electron dynamics:\\ Dynamic structure factors}%

\author{Pontus~Svensson}
\email{Electronic address: pontus.svensson@physics.ox.ac.uk}
\affiliation{Department of Physics, University of Oxford, Oxford OX1 3PU, United Kingdom}%

\author{Yusuf~Aziz}
\affiliation{AWE, Aldermaston, Reading, Berkshire RG7 4PR, United Kingdom}%

\author{Tobias~Dornheim}
\affiliation{Center for Advanced Systems Understanding (CASUS), D-02826 Görlitz, Germany}%
\affiliation{Helmholtz-Zentrum Dresden-Rossendorf (HZDR), D-01328 Dresden, Germany}

\author{Sam~Azadi}
\affiliation{Department of Physics, University of Oxford, Oxford OX1 3PU, United Kingdom}%

\author{Patrick~Hollebon}
\affiliation{AWE, Aldermaston, Reading, Berkshire RG7 4PR, United Kingdom}%

\author{Amy~Skelt}
\affiliation{AWE, Aldermaston, Reading, Berkshire RG7 4PR, United Kingdom}%

\author{Sam~M.~Vinko}
\affiliation{Department of Physics, University of Oxford, Oxford OX1 3PU, United Kingdom}%
\affiliation{Central Laser Facility, STFC Rutherford Appleton Laboratory, Didcot OX11 0QX, United Kingdom}%

\author{Gianluca~Gregori}
\affiliation{Department of Physics, University of Oxford, Oxford OX1 3PU, United Kingdom}%

% \date{\today}%
\date{July 11, 2024}%

\begin{abstract}
We present two methods for computing the dynamic structure factor for warm dense hydrogen without invoking either the Born-Oppenheimer approximation or the Chihara decomposition, by employing a wave-packet description that resolves the electron dynamics during ion evolution. First, a semiclassical method is discussed, which is corrected based on known quantum constraints, and second, a direct computation of the density response function within the molecular dynamics. The wave packet models are compared to PIMC and DFT-MD for the static and low-frequency behaviour. For the high-frequency behaviour the models recover the expected behaviour in the limits of small and large momentum transfers and show the characteristic flattening of the plasmon dispersion for intermediate momentum transfers due to interactions, in agreement with commonly used models for x-ray Thomson scattering. By modelling the electrons and ions on an equal footing, both the ion and free electron part of the spectrum can now be treated within a single framework where we simultaneously resolve the ion-acoustic and plasmon mode, with a self-consistent description of collisions and screening.
\end{abstract}

\maketitle

%%%%%%%%%%%%%%%%%%%%% INTRODUCTION %%%%%%%%%%%%%%%%%%%%%
\section{Introduction}\label{sec:intro}
The scattering of light from massive particles is one of the primary diagnostics for many-body systems, where the change in direction and energy of the photon carries information about the momentum and energy states within the systems~\cite{sturm1993dynamic,crowley2013x,crowley2014quantum}. To first order, the nonresonant scattering of electrons~\footnote{In principle ions act as scatterers as well, but this is suppressed relative to the electron contribution by the electron-ion mass ratio, even if the ions affect the scattering due to their influence on the electrons.} is described by the $\vec{A}^2$ term in the interaction Hamiltonian~\cite{sakurai1967advanced,crowley2013x}, $\vec{A}$ being the vector potential of the electromagnetic field, resulting in a differential cross section proportional to the electron dynamic structure factor $S_{ee}(\vec{k}, \omega)$~\cite{sturm1993dynamic},
\begin{equation}
    \frac{d^2 \sigma}{d\Omega d\omega} \propto S_{ee}(\vec{k}, \omega),
\end{equation}
for an energy and momentum change of $\hbar\omega$ and $\hbar\vec{k}$, respectively. The dynamic structure factor is the Fourier transform of the two-time density-density correlation function,
\begin{equation}
    S_{\alpha\beta}(\vec{k}, \omega) = \frac{1}{2\pi} \int_{-\infty}^{\infty}\!\!d\tau\, \frac{e^{i \omega \tau}}{\sqrt{N_{\alpha} N_{\beta} }} \left\langle \Hat{n}^{\alpha}_{\vec{k}}(\tau) \Hat{n}^{\beta}_{-\vec{k}}(0) \right\rangle_{T},
    \label{eq:DSF_definition}
\end{equation}
where $\Hat{n}^{\alpha}_{\vec{k}}(\tau)$ is the \textit{Heisenberg operator} for the spatially Fourier transformed density operator of particles of type $\alpha$, $N_{\alpha}$ is the number of such particles and $\langle \,\cdot\, \rangle_{T}$ is the thermal and quantum mechanical average \cite{sturm1993dynamic}. Therefore, the scattering is related to density fluctuations on length and timescales of $|\vec{k}|^{-1}$ and $\omega^{-1}$ respectively. 

In this work, we consider dense plasma systems -- warm dense matter (WDM) -- where ions are strongly coupled and electrons are partially Fermi degenerate~\cite{bonitz2020ab}, essential for describing a variety of astrophysical conditions \cite{guillot1999interiors,lorenzen2014progress,militzer2021first,saumon1992role,chabrier1993quantum,chabrier2000cooling,haensel2007neutron,daligault2009electron} but now also achievable in the laboratory~\cite{falk2018experimental,frank2019exoplanets} with static compression~\cite{konopkova2016direct} and laser drive using both x-ray~\cite{levy2015creation,fletcher2015ultrabright} and optical probes~\cite{kritcher2008ultrafast,smith2018equation}. In particular, the fuel capsule in an inertial confinement fusion experiment~\cite{nuckolls1972laser,abu2022lawson} transitions through the WDM regime~\cite{hu2018review} where improved descriptions are needed to model experiments~\cite{hu2015impact}. 

The measurement we focus on here is x-ray Thomson scattering (XRTS)~\cite{glenzer2009x}. Certain properties can be extracted from the XRTS spectrum in a model-independent manner -- e.g., temperature via detailed balance~\cite{glenzer2007oobservations,doppner2009temperature,descamps2020approach,dornheim2022accurate,dornheim2023imaginary} -- however, more commonly a comparison with a synthetic spectrum is used for the analysis~\cite{glenzer2007oobservations,lee2009x,ma2013x,kasim2019inverse}. Most XRTS analyses rely on the Chihara decomposition~\cite{glenzer2009x,falk2018experimental}, where $Z_B$ and $Z$ electrons per ion are labelled as bound and free, respectively, yielding the combined structure factor~\cite{chihara1987difference,chihara2000interaction},
\begin{equation}
    \begin{aligned}
        S_{ee}(\vec{k}, \omega) = \left|f(\vec{k}) + n(\vec{k}) \right|^2 &S_{ii}(\vec{k}, \omega)\\
        + Z &S_{ee}^0(\vec{k}, \omega) + Z_B S_{bf}(\vec{k}, \omega).
    \end{aligned}
    \label{eq:chihara}
\end{equation}
In the above, the ion-ion structure factor $S_{ii}$ is scaled by the form factors of the bound $f(\vec{k})$ and screening $n(\vec{k})$ electron cloud representing the adiabatic electron motion following the ions, $S_{ee}^{0}$ represents the correlations in the free-electron population, and the bound-free term $ S_{bf}$ arises from fluctuations in the bound-state populations. The spectrum includes a rich set of physics with dynamics on both the ionic and electronic timescales, however, uncertainties arise when separate models are used for all the terms~\cite{baczewski2016x} and the distinction between bound and free states can be problematic under WDM conditions~\cite{gawne2023investigating}.

The ionic structure can be modelled by the \textit{hypernetted chain} (HNC) approximation \cite{gregori2007derivation,glenzer2009x,hansen1993theory}, but the state-of-the-art models are Monte Carlo (MC) or molecular dynamics (MD) which explicitly treat the many-body structure and strong correlations. The interaction in such explicit models can be based on effective ion interactions \cite{ichimaru1982strongly,mithen2011extent,moldabekov2018structural,kahlert2020thermodynamic}, \textit{density functional theory} (DFT) using orbital-free \cite{lambert2007properties,white2013orbital} or Kohn-Sham \cite{collins1995quantum,ruter2014ab} descriptions, or \textit{path integral Monte Carlo} (PIMC) simulations \cite{dornheim2022effective,dornheim2023matter}. For the methods that solve for the electron density, the screening cloud $n(\vec{k})$ can be computed self-consistently~\cite{plagemann2015ab}.

The base model for the free electron contribution $S_{ee}^0$ is based on the random phase approximation (RPA) \cite{lifshitz2017course}, where electron-ion collisions are treated by the Mermin formulation of the dielectric function \cite{ropke1999lindhard,selchow2001dynamic,holl2007thomson,thiele2008plasmon,plagemann2012dynamic} and electron correlations using \textit{local-field corrections} (LFC)~\cite{wierling2009dynamic,fortmann2010influence}. To model $S_{ee}^{0}(\vec{k}, \omega)$ an explicit treatment of the electron state is needed~\cite{vorberger2015ab} and the electron time evolution should be resolved to model the $\omega$-dependence. \textit{Time dependent DFT} (TD-DFT)~\cite{magyar2016stopping} models time-varying electron dynamics and $S_{ee}^0$~\cite{baczewski2016x,frydrych2020demonstration}, however, it is limited to electron timescales. Alternatively, \textit{linear response TD-DFT} can be formulated without explicit time evolution~\cite{ramakrishna2019ab,ramakrishna2021first,schorner2023x} commonly performed in the adiabatic approximation~\cite{moldabekov2023linear}. Statistical methods, e.g., PIMC, exactly treat the full electron-ion dynamics beyond the Born-Oppenheimer approximation, see the recent Ref.~\cite{dornheim2024ab}, but do not have direct access to the frequency resolved properties; instead reconstructions based on LFC's have been suggested for the warm dense electron gas~\cite{dornheim2018ab,groth2019ab,hamann2020dynamics}.

The bound electron properties are commonly taken from atomic data and computations~\cite{gregori2003theoretical,glenzer2009x}, and require a detailed description of the bound states. The bound-free term primarily contributes at large $\omega$ but can overlap with the free term~\cite{schumacher1975incoherent,gregori2003theoretical,glenzer2009x,mattern2013theoretical}.

It is therefore desirable to have a single MD framework with explicit electron dynamics to model $S_{ee}^{0}$ while still being able to model ion dynamics $S_{ii}$ simultaneously, for system sizes large enough to describe collective phenomena. Semiclassical approaches of two-component plasma \cite{minoo1981temperature,hansen1983thermal,filinov2004temperature,glosli2008molecular} can be used as a starting point to bridge the timescale gap. Limited quantum effects are introduced using wave-packet molecular dynamics (WPMD), which constrains the electron wave function to a sub-manifold in Hilbert space \cite{feldmeier2000molecular,lasser2022various}, allowing for the time evolution of electronic properties over the characteristic timescale of ion dynamics. Both semiclassical~\cite{ortner1997quasiclassical,ebeling1998quasiclassical,selchow2001dynamic,golubnychiy2001dynamical,golubnychiy2002plasmon,zwicknagel2003dynamic} and WPMD~\cite{zwicknagel2006wpmd} electron models have been used to extract plasmon resonances. Our work will focus on the electron dynamics as well, using WPMD, but we will also simultaneously resolve the ion dynamics. A particular focus will be on the intrinsically quantum aspects of the scattering formulated from a real-time perspective, which is of substantial importance for modelling WDM conditions. 

The manuscript is structured as follows. Section \ref{sec:models} will describe the test system and the computational models, which are compared in sections \ref{sec:static_structure} and \ref{sec:classical_DSF} for static and dynamic properties, especially sections \ref{sec:classical_DSF} discusses the classical formulation for the dynamic structure factor. Quantum aspects are considered in section \ref{sec:quantum_DSF} and how impulse computations are used to go beyond the classical treatment in section \ref{sec:impulse_computations}. Frequency resolved properties are compared with commonly employed models in section \ref{sec:DSF_comparison}. A concluding summary is provided in section \ref{sec:conclusion}.

%%%%%%%%%%%%%%%%%%%%%%% METHOD %%%%%%%%%%%%%%%%%%%%%%%
\section{Simulation models}\label{sec:models}
\subsection{System parameters}
The primary test system under investigation is a homogeneous and isotropic sample of warm dense hydrogen with a density of $r_s = a_i / \bohr = (4\pi \bohr^3 n_i / 3)^{-1/3} = 2$ ($n_i = n_e \approx 2\times 10^{23}\,\text{cm}^{-3}$) and temperature $T = 250\,\text{kK} \approx 21.5\,\text{eV}$. The system has moderately coupled protons
\begin{equation}
    \Gamma_{pp} = \frac{Z^2 e^2}{4\pi\varepsilon_0 a_{i} \boltz T} \approx 0.63,
\end{equation}
and partially degenerate electrons
\begin{equation}
    \theta = \frac{\boltz T}{ E_{F} } \approx 1.72,
\end{equation}
where $e$ is the proton charge, $\varepsilon_0$ the vacuum permittivity, $\boltz T$ the temperature in energy units and the Fermi energy $E_{F} \approx 12.5\,\text{eV}$. The conditions are such that a wave packet description with pair-exchange interactions (see below) is appropriate. 

Under such conditions, hydrogen is expected to be ionised. The Saha equation with Debye-Hückel or ion-sphere~\cite{stewart1966lowering} IPD models predicts full ionisation, and all electrons are treated as free, i.e., $f(\vec{k}) = 0$, $Z = 1$ and $Z_B = 0$. The Chihara decomposition in equation \eqref{eq:chihara} therefore simplifies to
\begin{equation}
    S_{ee}(k, \omega) = |n(k)|^2 S_{ii}(k, \omega) + S_{ee}^{0}(k, \omega),
    \label{eq:chihara_fully_ionised}
\end{equation}
which only depends on the magnitude of the wave vector $k = |\vec{k}|$ and we average over equivalent $\vec{k}$ vectors in the simulations. The fermionic screening length~\cite{reinholz2005dielectric} and plasma frequency for the fully ionised system are $\lambda_s \approx 1.54\,\bohr \approx 0.81\,\text{\AA}$ and $\hbar\omega_p \approx 16.7\,\text{eV}$, respectively. 

\subsection{Wave-packet simulations}
The fundamental problem in resolving both electron and ion dynamics is the large proton to electron mass ratio, $m_p / m_e \approx 1836$, yielding significantly different timescales for the dynamics. The wave-packet approximation restricts the electron wave function to a parametrised functional form, allowing for a more rapid evaluation of the time evolution \cite{heller1975time,corbin1982semiclassical,feldmeier2000molecular,grabowski2014review}, where ionic dynamics can now be modelled~\cite{davis2020ion,angermeier2023disentangling}. Recently, an extension to the commonly used isotropic Gaussian states for wave packets was developed in terms of anisotropic Gaussians~\cite{svensson2023development}. This is what we use in this work.

For the temperature under consideration, a full Slater determinant~\cite{knaup2003wave,jakob2007wave,jakob2009wave} is not needed to treat the antisymmetrisation, but rather we employ an approximation in terms of Pauli interactions based on a pairwise antisymmetrisation \cite{klakow1994hydrogen,klakow1994semiclassical,su2007excited,su2009dynamics}, which allow for larger systems to be modelled. Angermeier \textit{et~al.}\ suggested an interaction between opposite spin electrons based on a spatially symmetric state~\cite{angermeier2021investigation}, however, we consider only half of the opposite spin interactions to have such form while the remainder interacts with the standard Pauli form -- along the lines of the spin-dependent interactions of Ref.~\cite{filinov2004temperature} -- by dividing each electron species into two groups.

Last, wave packets tend to expand indefinitely at sufficiently high temperatures, a common problem in WPMD simulations~\cite{knaup1999wave,knaup2001wave,knaup2003wave,ebeling2006method,morozov2009localization,davis2020ion}. An additional potential $\Hat{V}_{\Sigma}$ quadratic in the width, is employed to limit the expansion of electrons,
\begin{equation}
    \Hat{V}_{\Sigma} = \frac{A}{2} \sum_{i = 1}^{N_e} \left( \Hat{\vec{x}}_i - \left\langle \Hat{\vec{x}}_i \right\rangle \right)^2,
\end{equation}
where $\Hat{\vec{x}}_i$ is the position operator of particle $i$ and $A$ set to achieve the desired equilibrium width. In the main text ${A = 3\,\text{Ha}/\bohr^2}$ was used, but in appendix \ref{app:confinment} an additional confinement strength is tested. It is demonstrated that the ion structure is insensitive to the choice and only a minor dependence is seen for the electrons. 

For the computation, a system of $N_p = 2048$ protons and an equal number of electrons are used and evolved in time with a time step of $4 \times 10^{-4}\,\text{fs}$ to have good energy conservation. The system was initially thermalised with a simple velocity re-scaling thermostat to achieve the desired temperature -- see Ref.~\cite{svensson2023development} -- followed by simulations in the NVE ensemble. No data were collected during thermalisation and shorter thermalisations were interleaved between data collection segments. The presented data have been averaged over $70$ simulations of length $40\,\text{fs}$, except for impulse computations discussed in section~\ref{sec:impulse_computations}, computational details of which are described in appendix~\ref{app:impulse_linearity}.

\subsection{PIMC simulations}
The \textit{ab initio} PIMC method~\cite{ceperley1995path} is based on a stochastic evaluation of the thermal density matrix in coordinate representation. Most importantly, PIMC is, in principle, capable of giving quasiexact (i.e., exact within the given statistical error bars) results for a host of observables without the need for any external input. For fermions, such as the electrons in warm dense hydrogen, an additional obstacle is given by the notorious fermion sign problem~\cite{dornheim2019fermion,troyer2005computational}, which leads to an exponential increase in the required compute time with respect to parameters such as the system size, or towards low temperatures.

In this situation, it is a common practice to invoke the \textit{fixed-node approximation}~\cite{ceperley1991fermion}, which removes the sign problem at the cost of an approximation. Here, we follow the alternative strategy described in Ref.~\cite{dornheim2024ab} and carry out direct PIMC simulations of $N=32$ protons, where the sign problem is tamed by averaging over a large number of independent calculations. 

A detailed overview that includes the comparison of various PIMC approaches has recently been presented by Bonitz \textit{et al.}~\cite{bonitz2024principles}.

\subsection{DFT-MD simulations}
The test system was also modelled by DFT molecular dynamics (DFT-MD) in which the interatomic forces are calculated by DFT, using the CP2K package~\cite{kuhne2020CP2K}. The DFT based calculation is conducted using the mixed Gaussian and plane basis set, where the Kohn-Sham orbitals are expanded in contracted Gaussians, whereas the electronic charge density is represented using plane waves. For the former, we use an accurate triple basis set with additional sets of polarization functions TZV2PX~\cite{vandeVondele2007gaussian}, while for the latter we employ a cutoff of $500\,\text{Ha}$. The local density approximation (LDA) is used as the exchange-correlation functional with norm-conserving pseudopotential~\cite{goedecker1996seperable}. Only the $\Gamma$-point is used for the Brillouin zone integration.

The molecular dynamics was performed with $128$ protons in a cubic box with periodic boundary conditions. An ionic time step of $0.0725\,\text{fs}$ was seen to conserve the appropriate energy~\cite{wentzcovitch1992energy}. The system was thermalised using a Nosé-Hoover thermostat with a time constant of $100\,\text{fs}$ after which the statistics are accumulated using the NVE ensemble for $1700\,\text{fs}$.

%%%%%%%%%%%%%%%%%%%% Static structure %%%%%%%%%%%%%%%%%%%%
\section{Static and ionic properties}\label{sec:static_structure}
\begin{figure*}
    \centering
    %\subfloat{\includegraphics[width=0.33\linewidth,trim={0 0 1.30cm 0},clip]{Figures/static_structure_paper.jpg}}
    %\subfloat{\includegraphics[width=0.33\linewidth,trim={0 0 1.30cm 0},clip]{Figures/screening_paper.jpg}}
    %\subfloat{\includegraphics[width=0.33\linewidth,trim={0 0 1.30cm 0},clip]{Figures/ion_dispersion_paper.jpg}}
    \includegraphics[width=\linewidth]{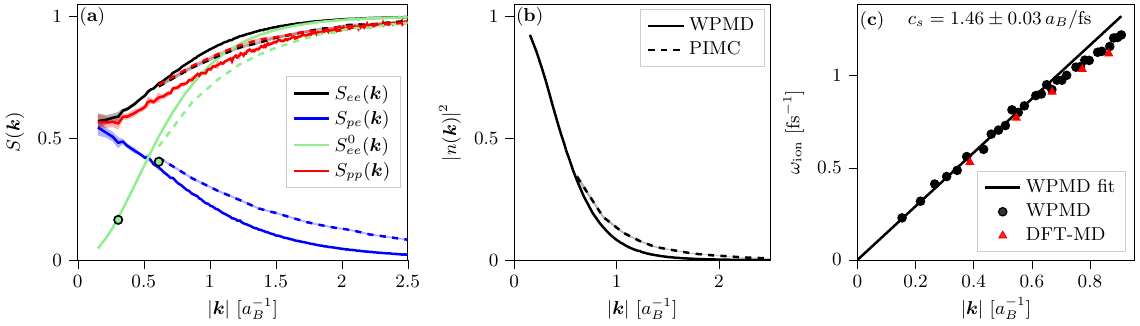}
    \caption{Comparisons of static properties and ion dynamics between the WPMD (solid) and PIMC (dashed).  Shaded regions correspond to $95\%$ confidence intervals estimated from different initial configurations. (a) Static structure factors for electron-electron $S_{ee}(\vec{k})$, proton-proton $S_{pp}(\vec{k})$, proton-electron $S_{pe}(\vec{k})$ and 'free' electron-electron $S_{ee}^{0}(\vec{k})$ where the ion screening cloud has been subtracted. Dots show the static structure computed within the impulse response formulation, see sections \ref{sec:impulse_computations} and \ref{sec:DSF_comparison}. (b) The screening form factors $n(\vec{k})$ within the different models. Errors within each model are comparable to line width. (c) Ion acoustic dispersion within WPMD (circles) and DFT-MD (triangles) based on a constrained GCM-fit \cite{wax2013effective,schorner2022extending}, and linear fit $\omega_{\text{ion}}(\vec{k}) = c_s |\vec{k}|$ for ion sound speed $c_s$, fitted for $|\vec{k}| \leq 0.5\, \bohr^{-1}$ in the wave packet model.}
    \label{fig:reference_data}
\end{figure*}
Before considering the time-dependent properties of the electrons, associated with inelastic scattering, the static properties and the ion dynamics are compared within the different computational models. The structural correlations are characterised by the static structure factor, which is the zero time separation of the intermediate scattering function, i.e., $S_{\alpha\beta}(\vec{k}) = F_{\alpha\beta}(\vec{k}, 0)$. The classical intermediate scattering function,
\begin{equation}
    F_{\alpha\beta}(\vec{k}, \tau) =  \frac{1}{\sqrt{N_{\alpha}N_{\beta}}} \sum_{i, j} \left\langle e^{-i \vec{k} \cdot \left( \vec{r}_i(t+\tau) - \vec{r}_j(t) \right)} \right\rangle_{t}, 
    \label{eq:classical_ISF}
\end{equation}
is computed based on the center of mass positions $\vec{r}_i(t)$ ($\vec{r}_j(t)$) at time $t$ for all $N_\alpha$ ($N_\beta$) particles of type $\alpha$ ($\beta$). The time average $\langle \cdot \rangle_t$ replaces the thermal average in the MD simulations. Figure~\ref{fig:reference_data}(a) shows the static structure factors, $S_{\alpha\beta}(\vec{k})$, for WPMD and PIMC. The wave-packet model is seen to agree well with the PIMC except for a reduced electron structure for large $\vec{k}$, due to the finite size of wave packets which softens the short-range interaction and lowers the density fluctuations below their typical size. However, the wave packet predicts more structured ions for intermediate $\vec{k}$ values interpreted as a weaker screening which is supported by Fig.~\ref{fig:reference_data}(b). This is likely due to the limited functional form for the electron density, even if the ionic structure was shown to be rather insensitive to the choice of confinement potential (see appendix \ref{app:confinment}). The ion structure in the DFT-MD agrees well with the PIMC results.

Figure~\ref{fig:reference_data}(b) isolates the form factors of the screening cloud within the different models from the cross-species structure factors, $|n(\vec{k})| = |S_{ie}(\vec{k}) / S_{ii}(\vec{k})|$, equivalent to how it has been done for DFT based methods in real-space~\cite{plagemann2015ab}. The form factor of the screening cloud in PIMC is observed to be larger than in WPMD for large $\vec{k}$, which is once more attributed to the restricted form of the wave packet not allowing gradients below the typical size of the packets.

Both WPMD and PIMC have access to full electron structure factors and the ``free'' contribution can be extracted, ${S_{ee}^{0}(\vec{k}) = S_{ee}(\vec{k}) - |n(\vec{k})|^2 S_{ii}(\vec{k})}$, not directly possible in a standard Born-Oppenheimer formulations~\cite{vorberger2015ab}. The ``free'' structure factors tend to agree for wavelengths larger than the wave packet size. Direct access to the ``free''-electron contribution allows for the study of the electron dynamics and high-frequency modes, which will be the main topic in sections \ref{sec:classical_DSF} -- \ref{sec:DSF_comparison}. However, first we will consider the low-frequency ion-acoustic modes which require a larger number of particles to model.

The WPMD and DFT-MD can access sufficiently long length scales (number of ions $N_i$) and timescales (physical time simulated) to model ion-acoustic waves. However, the required time step is substantially different as the electron dynamics are not modelled in DFT-MD. The intermediate scattering function is computed according to equation \eqref{eq:classical_ISF} for protons ($\alpha = \beta = p$) and the result is fitted with a generalised collective mode (GCM) approach with one diffusive mode and one oscillating mode, constrained by the first three (classical) sum rules \cite{wax2013effective,schorner2022extending}. This parameterisation has three independent parameters for each $\vec{k}$, one of which -- the oscillation frequency $\omega_{\text{ion}}$ of the oscillating mode -- is shown in Fig.~\ref{fig:reference_data}(c). From the clear linear trend for small $\vec{k}$ the ionic sound speed is measured \cite{hansen1993theory,mithen2011extent,kahlert2020thermodynamic,schorner2022extending} in the wave-packet model to be $c_s = 1.46\,\bohr/\text{fs} \approx 77.3\,\text{km}/\text{s}$. Comparing this to the DFT-MD data, good agreement for the dispersion is seen in Fig.~\ref{fig:reference_data}(c), both in regards to the linear part of the dispersion and where nonhydrodynamic effects are clearly seen in the dispersion, here around $\vec{k} \approx 0.7\,\bohr^{-1}$. The limited fitting window for linear dispersion makes any prediction of the ion-sound speed uncertain for the DFT-MD data and we refrain from such extraction.  

%%%%%%%%%%%%%%%%%%%% Classical DSF %%%%%%%%%%%%%%%%%%%%
\section{Classical dynamic structure factors}\label{sec:classical_DSF}
\begin{figure}
    \centering
    %\subfloat{\includegraphics[width=\linewidth]{Figures/intermidiet_scattering_paper.jpg}}\\
    %\subfloat{\includegraphics[width=\linewidth]{Figures/classcial_DSF_paper.jpg}}
    \includegraphics[width=\linewidth]{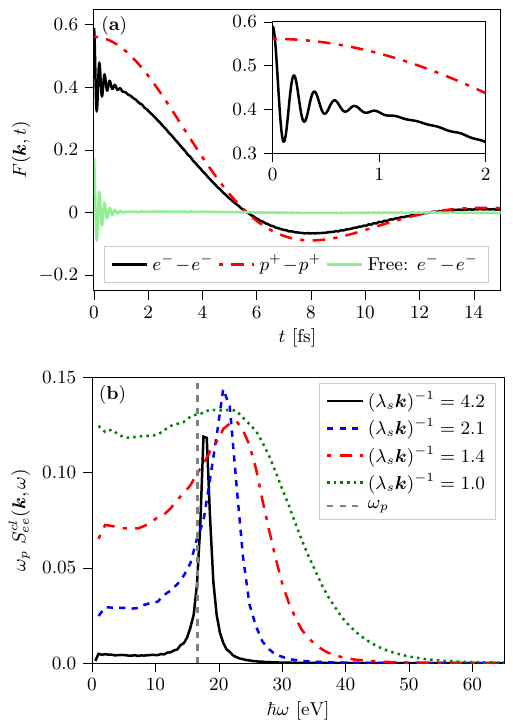}
    \caption{Classical (a) intermediate scattering functions $F(\vec{k}, t)$ at $k = 0.31\,\bohr^{-1}$ and (b) classical 'free'-electron dynamic structure factor $S_{ee}^{\text{cl}}(\vec{k}, \omega)$. (a) Electron dynamics ($e^- - e^-$) has two distinct features, the dynamics of the ion screening cloud set by the ion dynamics ($p^+ - p^+$) and the plasma oscillations. The latter is isolated by the subtraction $F_{ee}(\vec{k}, t) - |n(\vec{k})|^2 F_{pp}(\vec{k}, t)$ (Free: $e^- - e^-$). (b) The dynamic structure factor for 'free' electrons showing a clear plasmon feature in the collective regime ($(\lambda_s \vec{k})^{-1} \gg 1$) which is broadened for larger $\vec{k}$.}
    \label{fig:classical_DSF}
\end{figure}

The classical formulation of the intermediate scattering function, equation \eqref{eq:classical_ISF}, can be applied to the electrons as previously done in both classical MD \cite{ortner1997quasiclassical,ebeling1998quasiclassical,selchow2001dynamic,zwicknagel2003dynamic} and WPMD \cite{zwicknagel2006wpmd}; however, at the conditions under investigation the low electron inertia and the resulting high-frequency dynamics will be subject to quantum corrections -- qualitatively the quantum recoil and detailed balance -- something which will be discussed in the following section, and computations going beyond this approximation are discussed in section \ref{sec:impulse_computations}. Nevertheless, the classical formulation can clearly demonstrate the two-timescale nature of the dynamics, which is seen in Fig.~\ref{fig:classical_DSF}(a) for the WPMD. 

The longer timescale features in the electron dynamics are the correlations between the adiabatic screening clouds of the ions and therefore are related to the timescales of ion motion. This interpretation is supported by the Chihara decomposition and explicitly tested by isolating the ``free'' electron dynamics via subtraction of the screened ion dynamics,
\begin{equation}
    F_{ee}^{0}(\vec{k}, \tau) = F_{ee}(\vec{k}, \tau) - |n(\vec{k})|^2 F_{pp}(\vec{k}, \tau),
\end{equation}
which accounts for the long-time correlations to a high degree of accuracy. Computationally, this is seen before the ion dynamics have converged as the electrons will rapidly adjust to the quasiinstantaneous ionic configuration.

The remaining high-frequency oscillations in Fig.~\ref{fig:classical_DSF} describe the ``free'' electron dynamics, $S_{ee}^{0}$, shown in Fig.~\ref{fig:classical_DSF}(b), with a clear resonance (plasmon feature) in the collective regime $(\lambda_s \vec{k})^{-1} \gg 1$. This is broadened as we approach the noncollective behaviour $(\lambda_s \vec{k})^{-1} < 1$. The width of the plasmon feature for the more collective cases shown in Fig.~\ref{fig:classical_DSF}(b) is set primarily by the electron-ion collisions, which together with the dynamic screening, is self-consistently calculated in the WPMD.

In general, the data can be interpreted in the context of the Chihara decomposition. However, our dynamic WPMD models do not rely on such approximation, as the low-frequency modes are naturally added (see Fig.~\ref{fig:combined_spectrum}) to form the complete spectrum of excitations. The dynamics treat electrons and ions on an equal footing, nevertheless, separating the two behaviours is still beneficial for discussions and numerical comparisons. 

%%%%%%%%%%%%%%%% Quantum aspects of DSF %%%%%%%%%%%%%%%%%
\section{Quantum aspects of dynamic structure factors}\label{sec:quantum_DSF}
\begin{figure}
    \centering
    \includegraphics[width=\linewidth]{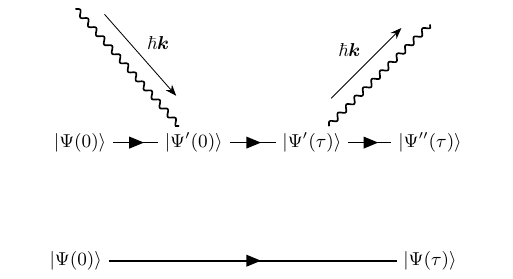}
    \caption{Schematic of the scattering mechanics which is implicit in the definition of the dynamic structure factor. Notation for the states are as given in the main text, and equation \eqref{eq:quantum_intermidiet_scattering} is a sum over terms $\bra{\Psi(\tau)}\ket{\Psi''(\tau)}$ where the absorption and emission are performed by different orbitals.}
    \label{fig:scattering}
\end{figure}

The dynamic structure factors in section \ref{sec:classical_DSF} are classical correlation functions, as the distributed nature of the electron state has been ignored. More importantly, the explicit time dependence of the Heisenberg operators must be considered. Neglecting this will lead to violation of  known constraints such as the $f$-sum rule and detailed balance~\cite{giuliani2008quantum,pines2018theory,chen2007interactions} in the classical formulation. In the Schrödinger picture, the time dependence stems from additional time propagators in the intermediate scattering function,
\begin{equation}
    \begin{aligned}
        &F_{\alpha\alpha}(\vec{k}, \tau) = \frac{1}{N_{\alpha}} \left\langle \bra{\Psi(\tau)} \Hat{n}_{\vec{k}}^{\alpha} e^{-i \Hat{H}\tau/\hbar} \Hat{n}_{-\vec{k}}^{\alpha} \ket{\Psi(0)}   \right\rangle_t\\
        &= \frac{1}{N_{\alpha}} \sum_{i,j}  \left\langle \bra{\Psi(\tau)} e^{-i\vec{k}\cdot \Hat{\vec{r}}_i} e^{-i \Hat{H}\tau/\hbar} e^{i \vec{k} \cdot \Hat{\vec{r}}_j} \ket{\Psi(0)}   \right\rangle_t,
    \end{aligned}
    \label{eq:quantum_intermidiet_scattering}
\end{equation}
where $\ket{\Psi(\tau)}$ is the simulation state at some time $t+\tau$. Note that the time propagator does not act directly on the simulation state but rather on the modified state,
\begin{equation}
    \ket{\Psi'} = e^{i \vec{k}\cdot\hat{\vec{r}}_j}\ket{\Psi(0)},
    \label{eq:modified_state}
\end{equation}
which carries a momentum shift of $\hbar\vec{k}$ compared to the original state. Equilibrium methods typically do not evolve the modified state $\ket{\Psi'}$; however, methods that can represent and propagate $\ket{\Psi'}$ can evaluate \eqref{eq:quantum_intermidiet_scattering} without additional assumptions. This idea extends to two-time correlation functions in general~\cite{breuer1997stochastic}. Equation \eqref{eq:quantum_intermidiet_scattering} can be interpreted as the overlap between the original state propagated in time $\ket{\Psi(\tau)}$ and the state $\ket{\Psi''(\tau)}$ which is created from the time propagation of the modified state for some time $\tau$ and the subsequent emission of the momentum $\hbar\vec{k}$. This process is schematically shown in Fig.~\ref{fig:scattering} and gives a clear picture of the scattering dynamics if a photon carries the momentum $\hbar\vec{k}$. The classical approximation discussed above therefore amounts to assuming this momentum shift does not influence the dynamics, arguably appropriate for ion dynamics in warm dense hydrogen which satisfy $\beta\hbar\omega_{\text{ion}} \ll 1$. However, the same cannot be said for the electrons, where for our specific system $\beta\hbar\omega_p \approx 0.8$.

To demonstrate the effect of the time evolution of the modified state, consider the noninteracting single-particle limit, appropriate for short-time dynamics and the $\vec{k} \rightarrow \infty$ limit. By writing the state in terms of plane-waves $\ket{\vec{p}}$, the self-contribution of the intermediate scattering function is
\begin{equation}
    \begin{aligned}
        F_{\alpha\alpha}(\vec{k}, \tau) &= \sum_{\vec{p},\vec{p'}} \bra{\vec{p}}\ket{\Psi(0)}\bra{\Psi(0)}\ket{\vec{p}'} e^{i\omega_{\vec{p}}\tau}\\
        &\hspace{1.2cm}\times\bra{\vec{p}'} e^{-i\vec{k}\cdot \Hat{\vec{r}}} e^{-i \Hat{H}\tau/\hbar} e^{i \vec{k} \cdot \Hat{\vec{r}}} \ket{\vec{p}}\\
        &= \sum_{\vec{p}} |\bra{\Psi(0)}\ket{\vec{p}}|^2 \;  e^{-i \tau \left( \omega_{\vec{p}+\hbar\vec{k}} - \omega_{\vec{p}} \right)},
    \end{aligned}
    \label{eq:single_particle_correlations}
\end{equation}
and $\hbar\omega_{\vec{p}} = \vec{p}^2/(2m)$ is the energy of the plane-wave state. The exponent now represents an interference effect between the unmodified and the modified states due to their difference in energy. From expanding the energy difference $\omega_{\vec{p}+\hbar\vec{k}} - \omega_{\vec{p}} = (\vec{p} \cdot \vec{k})/m + \omega_{R}$ we note two distinct contributions. The first term is translation for short times, which is captured in the classical formulation \eqref{eq:classical_ISF}, while the second term is the recoil energy $\hbar\omega_R = \hbar^2\vec{k}^2/(2m)$, the associated energy of the momentum kick, and missing in the classical formulation. On thermal averaging with a Maxwell-Boltzmann distribution, the resulting dynamic structure factor is
\begin{equation}
    S_{\alpha\alpha}(\vec{k}, \omega) = \sqrt{\frac{m\beta}{2\pi \vec{k}^2}} e^{-\frac{m\beta}{2\vec{k}^2}(\omega - \omega_R)^2},
    \label{eq:single_particel_limit}
\end{equation}
now including the Compton shift and satisfies the $f$-sum rule $\langle \omega \rangle_S = \omega_R$, where $\langle \omega^n \rangle_{S}$ is the $n$th moment of $S(\vec{k}, \omega)$. The same result can be derived using operator notation and commutators; see, e.g., Ref.~\cite{scopigno2005microscopic}.

It is worth mentioning that this type of shift in the resonance by $\omega_R$ in equation \eqref{eq:single_particel_limit} due to the quantum recoil is commonly added in the kinetic formulations for the evaluation of resonances~\cite{melrose2008quantum} and in particular the resonance of collective plasma oscillations $\omega_0$ is shifted like~\cite{thiele2008plasmon,melrose2010dispersion}
\begin{equation}
    \omega_0^2 \rightarrow \omega_0^2+\omega_R^2.
    \label{eq:collective_recoil}
\end{equation}

A second nonclassical feature of the dynamic structure factor is detailed balance, 
\begin{equation}
    S(\vec{k}, -\omega) = e^{-\beta\hbar\omega} S(-\vec{k}, \omega),
\end{equation}
a consequence of the occupation of states in the statistical formulation~\cite{giuliani2008quantum,pines2018theory}. This property is not retained in the classical formulation due to time inversion symmetry $F_{\alpha\beta}(\vec{k}, \tau) = F_{\alpha\beta}(\vec{k}, -\tau)$ where positive and negative frequencies have equal amplitude. The complete symmetry, $F_{\alpha\beta}(\vec{k}, \tau) = F^{*}_{\alpha\beta}(\vec{k}, -\tau)$, only guarantees real structure factors. Ortner \textit{et~al.}~\cite{ortner1997quasiclassical} suggested that molecular dynamics computes a classical version of the fluctuation-dissipation theorem while still providing a good description of the dielectric response, an argument that results in the traditional correction formula for detailed balance:
\begin{equation}
    S(\vec{k}, \omega) = \frac{\beta\hbar\omega}{1 - \exp\left(-\beta\hbar\omega\right)} S^{\text{cl}}(\vec{k}, \omega)
    \label{eq:DB_correction_classical}
\end{equation}
where we have added the superscript "cl" to denote the classical structure factor. The same expressions can be arrived at by making connections to the relaxation function~\cite{lovesey1984theory}. Note that the above correction respects the $f$-sum rule $\langle \omega \rangle_{S} = \beta \hbar \langle \omega^2 \rangle_{S^{\text{cl}}} / 2$, assuming the classical result $\langle \omega^2 \rangle_{S^{\text{cl}}} = \boltz T \vec{k}^2 / m$, but it does change the static structure factor. This results in large errors for $\beta\hbar\omega \gg 1$. 

Despite the widespread use of equation \eqref{eq:DB_correction_classical}, especially for ion dynamics \cite{baron2020introduction,scopigno2005microscopic,gregori2009low}, it cannot be employed blindly for all corrections. This is exemplified by expanding equation \eqref{eq:single_particel_limit},
\begin{equation}
    S(\vec{k}, \omega) = e^{\frac{\beta\hbar}{2} \left(\omega - \omega_R/2 \right)} S^{\text{cl.}}(\vec{k}, \omega),
    \label{eq:DB_correction_single_particel}
\end{equation}
which is very different from equation \eqref{eq:DB_correction_classical} for $\beta\hbar\omega \gtrsim 1$ and the traditional correction formula cannot be used in the single-particle limit. Therefore we propose an interpolation between the two cases, whereby isolating the collective and single-particle modes each part is corrected according to equation \eqref{eq:DB_correction_classical} -- where we incorporate the shift due to the recoil in the collective mode -- and \eqref{eq:DB_correction_single_particel}, respectively. The different parts of the spectrum are isolated by fitting a combined GCM (one diffusive and one collective mode) with a weighted single particle spectrum where the constraints from the first three sum-rules were adjusted appropriately. Further details are given in appendix \ref{app:DSF_detailes}. The formulation will satisfy the $f$-sum rule by construction. This method is demonstrated and compared in section \ref{sec:DSF_comparison} with explicit computation of the density response function, described in the following section.

Last, we need to point out that the benefit of deriving the electron dynamics from equations \eqref{eq:quantum_intermidiet_scattering} and \eqref{eq:modified_state}, rather than via the density response formulation, is the extension to nonequilibrium systems since they do not rely on the fluctuation-dissipation theorem. A nonequilibrium formulation would add to the discussion on the influence of gradients and nonthermal distribution functions in XRTS -- extensively discussed recently~\cite{fortmann2009thomson,kozlowski2016theory,belyi2018thomson,belyi2018theory,beuermann2019thomson,vorberger2024revealing} -- however, the wave-packet approximation, and especially the effect of the confinement potentials, strongly influences the result. To illustrate this point, consider straight line motion as in equation \eqref{eq:single_particle_correlations}, where an additional suppression factor would be present when the modified $\bra{\Psi(\tau)}\Hat{\vec{r}}_i\ket{\Psi(\tau)}$ and unmodified particle $\bra{\Psi'(\tau)} \Hat{\vec{r}}_i \ket{\Psi'(\tau)}$ trajectories have separated a comparable distance to the wave-packet size, due to the restricted expansion of the wave packet imposed by the confinement potential. The density response formulation only considers the total electron density and is therefore less affected by these constraints. 
 
%%%%%%%%%%%%%%%%% Impulse simulations %%%%%%%%%%%%%%%%%%
\section{Density response coefficients and dynamic structure factors}\label{sec:impulse_computations}
An alternative route for computing the dynamic structure factors instead of via the two-time correlation functions $F_{\alpha\beta}(\vec{k}, \tau)$ is the computation of the \textit{density response function} $\chi(\vec{r}, \vec{r}', t - t')$ directly in the MD by the explicit introduction of an external potential similar to how it has previously been done with TD-DFT~\cite{baczewski2016x}. 

Within linear theory, Ichimaru \textit{et al.}~\cite{ichimaru1985theory} writes the density response for a two-component plasma due to an external perturbation $V_{\alpha}^{\text{ext}}$ acting on species $\alpha$ as 
\begin{equation}
    \begin{aligned}
        \delta &n_{\alpha}(\vec{r}, t) =\\
        &\sum_{\beta} \int_{-\infty}^{t}\!\!\! dt' \int\! d^3\vec{r}'\; \chi_{\alpha\beta}(\vec{r}, \vec{r}', t - t') V_{\beta}^{\text{ext}}(\vec{r}', t'),
    \end{aligned}
    \label{eq:density_resposne}
\end{equation}
where $\delta n_{\alpha}(\vec{r}, t)$ is the density response of species $\alpha$, $\chi_{\alpha\beta}$ are the appropriate response coefficients and the indices run over species, here $\alpha, \beta \in \{e, p\}$. The fluctuation-dissipation theorem now provides the remaining connection to the dynamic structure factor~\cite{ichimaru1985theory,sturm1993dynamic}
\begin{equation}
    S_{ee}(\vec{k}, \omega) = -\frac{\hbar}{1 - \exp\left(-\beta\hbar\omega\right)} \frac{\Im\left\{ \chi_{ee}(\vec{k}, -\vec{k}, \omega) \right\}}{\pi N},
    \label{eq:fluctuation-dissepation}
\end{equation}
where the response function has been Fourier transformed in space and time
\begin{equation}
    \begin{aligned}
        \chi_{\alpha\beta}(&\vec{k}, -\vec{k}, \omega)\\
        &= \int\! dt\!\! \int\! d^3\vec{r}\, d^3\vec{r}'\; e^{i\omega t - i \vec{k}\cdot (\vec{r} - \vec{r}')} \chi_{\alpha\beta}(\vec{r}, \vec{r}', t).
    \end{aligned}
\end{equation}
Within this formulation $S_{ee}(\vec{k}, \omega)$ automatically satisfies detailed balance and the response functions generally adhere to the $f$-sum rule as this is a consequence of mass conservation; see Ref.~\cite{giuliani2008quantum}. In appendix \ref{app:f_sum_chi} we discuss a nuance of how the finite wave packet size influences the results when the wavelength of the external perturbation is comparable in size and in particular the $f$-sum rule.

\begin{figure}
    \centering
    \includegraphics[width=\linewidth]{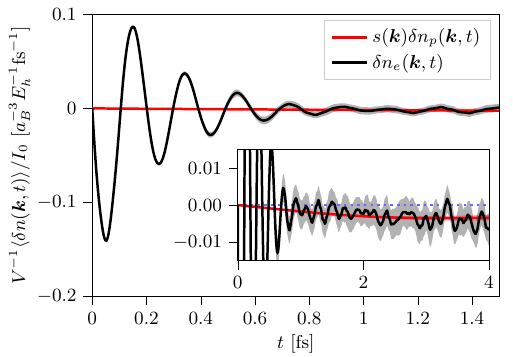}
    \caption{Fourier transform density response resolved in time for the electrons (perturbed species) and the protons at $\vec{k} = 0.31\,\boltz^{-1}$. The proton response is scaled with a screening factor $s(\vec{k})$ to match long-time electron response. The dominant signal is the high-frequency plasma oscillations onto a smaller amplitude proton motion on longer timescales (as shown in the inset). Subtraction of the proton signal isolates the high-frequency dynamics analogously to Fig.~\ref{fig:classical_DSF}. Shaded areas are 95$\%$ confidence intervals, estimated from variations between impulse simulations.}
    \label{fig:density_response}
\end{figure}

In investigating the electron dynamic structure factor, it is evident from equation \eqref{eq:density_resposne} that one should perturb the electrons and examine their response. Following the methodology of Sakko \textit{et~al.}~\cite{sakko2010time} a complex impulse of strength $I_0$ is introduced $V_{\alpha}^{\text{ext}}(\vec{r}, t) = I_0 \delta(t) \exp\left(i\vec{k}\cdot\vec{r}\right) \delta_{\alpha e}$ which in practice is modelled by two separate simulations. The density perturbations are averaged over multiple starting configurations, taken from a thermal simulation, to incorporate different ionic configurations and reduce the effect of thermal fluctuations. To further reduce the effect of the latter, the fluctuations in an unperturbed simulation starting from the same configuration can be subtracted, however, these vanish after proper averaging. An example of the resulting average density response, resolved in time, is shown in Fig.~\ref{fig:density_response}.

The impulse response formulation is primarily used to model the higher-frequency part of the spectrum and not the ion dynamics, as this limits the time after the impulse needs to be modelled. However, the protons are still treated fully dynamically and their density response is highlighted in the inset of Fig.~\ref{fig:density_response}. This response corresponds to the cross-species response function $\chi_{pe}$, not directly related to $\chi_{pp}$~\cite{ichimaru1985theory} which appears in the Chihara decomposition \cite{chihara1987difference}. However, a scaled proton response still explains the long-time (low-frequency) electron response and is subtracted to isolate the ``free'' electron dynamics. The low-frequency response is substituted by the higher resolution time data from the classical model in section~\ref{sec:classical_DSF} which behaves classically. Further details of the impulse simulations and the effect of a finite-strength impulse are provided in Appendix \ref{app:impulse_linearity}.

%%%%%%%%%%%%%%%%% Comparisons %%%%%%%%%%%%%%%%%%
\section{Dynamic structure factor: Comparison and predictions}\label{sec:DSF_comparison}%
\begin{figure*}
    \centering
    % \subfloat{\includegraphics[width=0.5\linewidth]{Figures/plasmon_paper.jpg}}
    % \subfloat{\includegraphics[width=0.5\linewidth]{Figures/plasmon_width_paper.jpg}}
    \includegraphics[width=\linewidth]{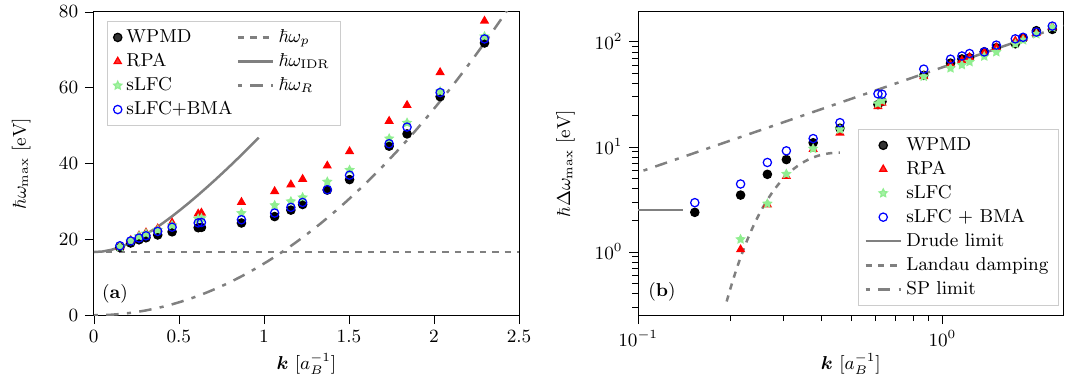}
    \caption{(a) Position and (b) width of the feature with the largest amplitude in the ``free'' electron part of the spectrum. Positions and widths (FWHM) obtained for Gaussian fit for detailed balance-corrected classical WPMD model (WPMD), the random phase approximation (RPA), with local field correction for UEG~\cite{dornheim2019static} (sLFC) and extended Mermin formulation (sLFC + BMA)~\cite{wierling2009dynamic,fortmann2010influence} with collision frequencies within the Born approximation~\cite{thiele2008plasmon} using ion structure from the wave-packet simulation. The analytical model for the limit $\vec{k} \rightarrow 0$ by Thiele \textit{et~al.}~\cite{thiele2008plasmon} labelled $\omega_{\text{IDR}}$ and the Drude limit of the Mermin formulation with collisions in BMA are shown. Model neglecting electron-ion collisions are compared to classical Landau damping rates in this limit~\cite{haas2011quantum}. The single particle limit (SP limit) is characterised by the recoil frequency $\hbar\omega_R$ and the width of the single-particle spectrum.}
    \label{fig:plasmon}
\end{figure*}%
Two computational models for computing the dynamic structure factor without the Born-Oppenheimer approximation using the wave-packet description have been presented. These models will now be compared to the ones commonly used for the interpretation of XRTS spectra discussed in the introduction.

In Fig.~\ref{fig:plasmon} the position and a measure of the width of the largest feature in the ``free'' electron spectrum are compared between a semiclassical formulation and analytical models. In the collective regime, $|\vec{k}| \leq \lambda_s^{-1}$, the spectrum is dominated by a clear plasmon feature whose position is well described by the analytical model of Thiele \textit{et~al.}\cite{thiele2008plasmon} when $|\vec{k}| < 0.5 \bohr^{-1}$. For larger $\vec{k}$ the wave packet model and RPA based models \cite{lifshitz2017course} predict a flattening of the plasmon dispersion in the transition from collective to noncollective scattering occurring at $\lambda_s |\vec{k}| = 1 - 3$. This effect is especially clear when electron interactions are included, exemplified here by the inclusion of static LFC (sLFC) parameterised for an \textit{uniform electron gas} by Dornheim \textit{et al.}~\cite{dornheim2019static}. The downshift compared to noninteracting RPA is arguably the beginning of the roton feature discussed for hydrogen at lower temperatures~\cite{dornheim2018ab,dornheim2022electronic,hamann2023prediction}. Furthermore, an additional downshift is seen when considering electron-ion collisions in Fig.~\ref{fig:plasmon}(a) in terms of the extended Mermin formulation~\cite{wierling2009dynamic,fortmann2010influence} where LFC's are treated as above and electron-ion collisions are modelled within the first Born approximation (sLFC+BMA)~\cite{thiele2008plasmon}. This type of downshift is commonly associated with an imaginary component of the dynamic collision frequency~\cite{fortmann2010influence}. The wave packet model generally predicts all of the above features of the resonance position, however, the downshifts seen are generally slightly larger compared to LCF+BMA. In the noncollective regime, for our test system $|\vec{k}| > 3 \lambda_s^{-1}$, all models converge to the single-particle limit, where the resonance position is set by the Compton shift $\omega_R$.

When considering the plasmon feature at small $\vec{k}$, the width is dominated by the electron-ion collisions and good agreement is seen between the wave-packet model and the sLFC+BMA, both of which are seen in Fig.~\ref{fig:plasmon}(b) to tend to a finite width similar to the Drude-limit of the Born-Mermin ansatz~\cite{fortmann2010influence}. The wave-packet model therefore predicts an electron-ion collision frequency similar to the first Born-approximation. At intermediate $\vec{k}$ the models which neglect electron-ion collisions achieve a similar width to the models which include the collisions (WPMD and sLFC+BMA), as Landau damping substantially contributes to the broadening of the plasmon feature, demonstrated by good agreement with the classical damping rate~\cite{haas2011quantum} in Fig.~\ref{fig:plasmon}(b). All models agree in the single particle-limit.

%\begin{figure}
%    \centering
%    \includegraphics[width=\linewidth,trim={0cm 0cm 0cm 2.5cm},clip]{Figures/comparison_real_paper_new.jpg}
%    \caption{Comparisons of 'free' dynamic structure factors in the collective regime $(\lambda_s \vec{k})^{-1} = 4.2,\; 2.1$, the intermediate region $(\lambda_s \vec{k})^{-1} = 1.4,\; 1.0$ and the non-collective single-particle limit $(\lambda_s \vec{k})^{-1} = 0.4$, within different models. The models presented are as discussed in figure \ref{fig:plasmon}.}
%     \label{fig:comparison_real}
% \end{figure}

\begin{figure}
    \centering
    \includegraphics[width=\linewidth]{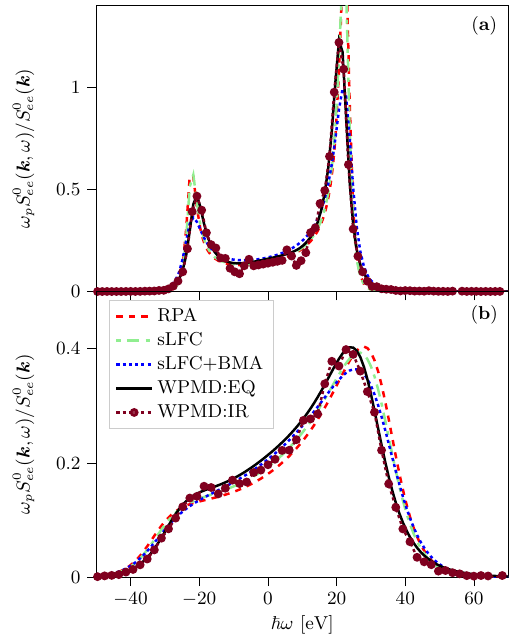}
    \caption{Comparisons of ``free'' dynamic structure factors in the collective regime $(\lambda_s \vec{k})^{-1} = 2.1$ (a) and $(\lambda_s \vec{k})^{-1} = 1.1$ (b). Models are as presented in Fig.~\ref{fig:plasmon} with the addition of the impulse response results (WPMD:IR) as described in section \ref{sec:impulse_computations}. A linear fit of $\chi_{ee}$ for small $\omega$ has been used to reduce noise. The impulse response agrees well with the semiclassical corrections for the WPMD (WPMD:EQ).}
    \label{fig:comparison_real}
\end{figure}

Some example spectra of the overall trends discussed in Fig.~\ref{fig:plasmon} are shown in Fig.~\ref{fig:comparison_real} along with the result of the impulse computations. For the conditions under consideration, $\beta\hbar\omega_p \approx 0.8$, the effect of detailed balance can be seen, demonstrating the need for a quantum description of the DSF. The result of the density response computations is seen to agree well with the semiclassical correction formula especially for the position and width of the resonance, marginal differences are seen at $\omega = 0$. The integral -- the static structure factor $S(\vec{k})$ -- differs between the models for larger $\vec{k}$, where the density response formulation yields a reduced $S(\vec{k})$. This is shown in Fig.~\ref{fig:reference_data}(a), where the semiclassical formula agrees well with the classical static structure. This is partially attributed to the finite size of the wave packet which explicitly enters in the impulse computations. The overall agreement between the formulations supports the semiclassical correction. All models presented predict similar spectral shapes, where the shifts and width of the resonance are the primary differences.   

\begin{figure}
    \centering
    \includegraphics[width=\linewidth]{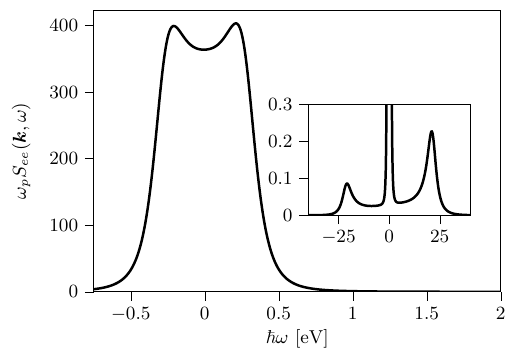}
    \caption{Complete electron structure factor incorporating both the ion-acoustic modes and the plasmon resonance at $\vec{k} = 0.31\,\bohr^{-1}$ for detailed balance corrected WPMD. The ionic timescale feature is narrow around $\omega = 0$ on top of the broader free electron contribution where substantial asymmetry between positive and negative frequencies can be seen. The ionic feature is fitted with a GCM fit (see section \ref{sec:classical_DSF}) and the electron feature is corrected for detailed balance according to section \ref{sec:quantum_DSF}.}
    \label{fig:combined_spectrum}
\end{figure}

Last, we demonstrate the ability of the two models to include ion dynamics along with the electron motion in Fig.~\ref{fig:combined_spectrum}, which shows the complete electron structure factors. The ion dynamics are processed by the fitting procedure discussed in section \ref{sec:static_structure} and a detailed balance correction according to equation \eqref{eq:DB_correction_classical} is applied as the ion dynamics satisfy $\beta\hbar\omega \ll 1$ (see Fig.~\ref{fig:combined_spectrum}). The ion data is scaled by the appropriate screening factor $|n(\vec{k})|^2$. The comparatively slow ion dynamics compared to plasma oscillations result in a narrow feature close to a $\delta$-function in a frequency range appropriate for the electron spectrum. When the ion feature is resolved, it has two distinct ion-acoustic features. However, the width of such acoustic modes does not allow for a clear separation of the diffusive mode which otherwise contributes roughly $52\%$ of the ionic feature.

%%%%%%%%%%%%%%%%% Conclusion %%%%%%%%%%%%%%%%%%
\section{Conclusion}\label{sec:conclusion}
X-ray Thomson scattering is one of the key diagnostics for warm dense matter conditions. It includes a detailed description of the microscopic dynamics of the plasma and is directly related to the electron structure factor. In particular, dynamics on both the electron (plasmon resonance) and ion (diffusion and ion-acoustic modes) timescales are present within the measurement. We have presented how real-time electron modelling with wave-packet molecular dynamics can be used to model features on both timescales within a single formulation with a self-consistent description of screening and collisions. The purely quantum aspects of the scattering have been discussed from a real-time modelling perspective.

Ultimately, two approaches have been presented to calculate the dynamic structure factor. The first is based on a semiclassical interpretation of the wave packet trajectories with detailed balance corrections and the second is based on explicit computation of the density response function within the MD, which is related to the scattering via the fluctuation-dissipation theorem. As a result of the separation of timescales, the ion and electron dynamics are treated partially separately -- as suggested by the Chihara decomposition -- for computational and interpretation reasons, where the quantum aspects of the electron dynamics must be accounted for. However, we go beyond the Born-Oppenheimer approximation as all the dynamics are modelled with electron and ion dynamics on an equal footing.

The presented models are compared with PIMC and DFT-MD for the static properties and ion dynamics, as well as with commonly used models for the interpretation of the XRTS spectrum. Similar static structures are observed in all models but some influence due to the wave-packet size is seen for wavelength comparable to the wave-packet size. Broadly the same spectral feature are seen in all models, with a narrow ionic feature incorporating a diffusive and ion-acoustic mode and the broader electron feature. The ion-acoustic mode dispersion is seen to agree with DFT-MD for a broad range of wavelengths. The plasmon feature shows a similar structure as RPA-based models when both electron interactions in terms of static local field corrections and electron-ion collisions are included. In particular, the wave packet models predict collisional broadening of the plasmon resonance similar to the first Born approximation while predicting a slightly larger downshift in the resonance position. 

The two models presented for the computation of the electron-electron dynamic structure factor agree well with each other, especially in terms of positions and widths of the resonances, where differences are seen in the static structure factor. This demonstrates that the semiclassical formulation corrects for the main quantum aspects.

The ability to model both electron and ion timescale phenomena within a single model allows for the treatment of most of the XRTS spectrum where bound state effects are the remaining missing feature. The modelling of bound states is beyond the limited functional form for the electron wave function used here, and the pairwise-exchange approximation.  

%%%%%%%%%%%%%%%%%%%%% BACKMATTER %%%%%%%%%%%%%%%%%%%%%
\begin{acknowledgments}
    We acknowledge stimulating discussions with T.~Campbell and D.~Plummer.
    The authors are gratefully for the use of computing resources provided at STFC Scientific Computing Department’s SCARF cluster, where the wave-packet computations have been performed.
    The PIMC calculations were performed at the Norddeutscher Verbund f\"ur Hoch- und H\"ochstleistungsrechnen (HLRN) under Grant No. mvp00024 and on the HoreKa supercomputer funded by the Ministry of Science, Research and the Arts Baden-W\"urttemberg and by the Federal Ministry of Education and Research.
    PS acknowledges funding from the Oxford Physics Endowment for Graduates (OXPEG). PS, SMV and GG acknowledge support from AWE-NST via the Oxford Centre for High Energy Density Science (OxCHEDS). SMV acknowledges support from the UK EPSRC Grant No. EP/W010097/1. TD: This work was partially supported by the Center for Advanced Systems Understanding (CASUS), financed by Germany’s Federal Ministry of Education and Research (BMBF) and the Saxon state government out of the State budget approved by the Saxon State Parliament. This work has received funding from the European Research Council (ERC) under the European Union’s Horizon 2022 research and innovation programme (Grant Agreement No. 101076233, "PREXTREME").  Views and opinions expressed are however those of the authors only and do not necessarily reflect those of the European Union or the European Research Council Executive Agency. Neither the European Union nor the granting authority can be held responsible for them.
\end{acknowledgments}

\appendix

%%%%%%%%%%%%%%%%% Confinment potential %%%%%%%%%%%%%%%%
\section{Confinement potential for wave packets}\label{app:confinment}
\begin{figure*}
    \centering
    %\subfloat{\includegraphics[width=0.50\linewidth,trim={0 0 0.0cm 0},clip]{Figures/static_structure_confinment.jpg}}
    %\subfloat{\includegraphics[width=0.50\linewidth,trim={0 0 0.0cm 0},clip]{Figures/classical_DSF_confinment.jpg}}
    \includegraphics[width=\linewidth]{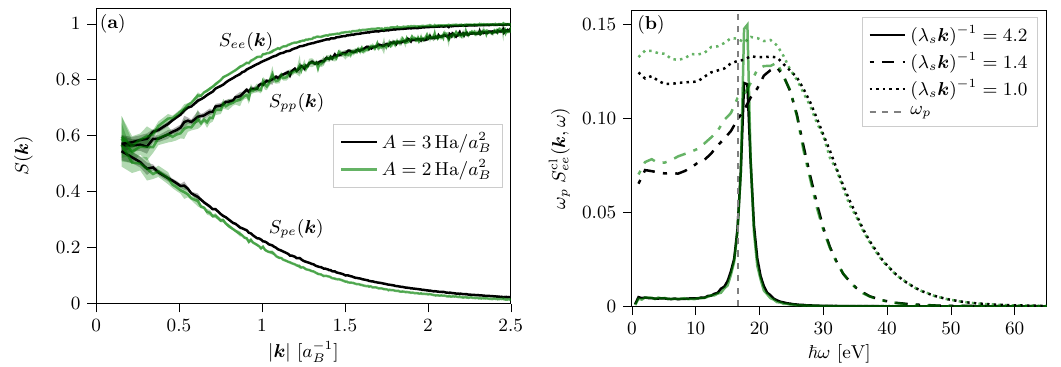}
    \caption{The effect of different confinement potential strengths for (a) static structure which is to be compared to Fig.~\ref{fig:reference_data}(a) and (b) the classical ``free'' electron dynamic structure factors, which are to be compared to Fig.~\ref{fig:classical_DSF}(b). The stronger confinment case has been averaged over $70$ data runs while the lower confinement case has been averaged over $25$, which is shown in the error estimates (shaded areas), which are $95\%$ confidence intervals estimated from the different independent runs.}
    \label{fig:confinment}
\end{figure*}%
The major unknown for the wave-packet model is how to set the confinement potential and which impact this regularisation has on the computations. A comprehensive attempt to answer such questions is beyond the scope of the current manuscript, but the effect on static structure factors $S(\vec{k})$ and classical ``free'' electron dynamic structure factors $S_{ee}^{0}(\vec{k}, \omega)$, for two different strengths are shown in Fig.~\ref{fig:confinment}.

The effect of the confinement potential is very minor on the ion structure but has some influence on the electron structure. In particular, the lower confinement case can be seen to give less structure in the electron-electron structure factor and a weaker ion screening, two results in agreement with having larger wave packet. However, the effect is overall minor. 

The general spectral shape of the classical dynamic structure factors is also shown in Fig.~\ref{fig:confinment}(b) to be rather insensitive to the choice of $A$, however, minor changes are seen due to variations in the collisionality due to different average packet sizes. This has been previously discussed in Ref.~\cite{morozov2009localization}. Overall, the results here are not too sensitive to the choice of $A$; however, it is a question which should be investigated further in general. 

%%%%%%%%%%%%%%%%%%%%% DSF Details %%%%%%%%%%%%%%%%%%%%%
\begin{widetext}
\section{Details on dynamic structure factor calculations}\label{app:DSF_detailes}
To smoothly interpolate between the collective and single particle detailed balance corrections of the dynamic structure factor, the classical structure factor is fitted by
\begin{equation}
    S^{\text{cl}}(\vec{k}, \omega) = S^{\text{GCM}}(\vec{k}, \omega) + C S^{\text{sp}}(\vec{k}, \omega),
\end{equation}
where
    \begin{equation}
         2\pi S^{\text{cl}}(\vec{k}, \omega) = 2B_0 \frac{\alpha}{\alpha^2 + \omega^2} + \frac{B_1\gamma}{\gamma^2 + (\omega_0 + \omega)^2} + \frac{B_1\gamma}{\gamma^2 + (\omega_0 - \omega)^2} + \frac{B_2 (\omega_0 + \omega)}{\gamma^2 + (\omega_0 + \omega)^2} + \frac{B_2 (\omega_0 - \omega)}{\gamma^2 + (\omega_0 - \omega)^2},
    \end{equation}
\end{widetext}
and
\begin{equation}
    S^{\text{sp}}(\vec{k}, \omega) = \sqrt{\frac{\beta m}{2\pi \vec{k}^2}} e^{-\frac{\beta m}{2 \vec{k}^2} \omega^2},
\end{equation}
are parameterised by $\vec{k}$-dependent fitting coefficients $\alpha, \omega_0, \gamma$ and $C$. The remaining coefficients are set by the first three classical sum rules
\begin{subequations}
    \begin{align}
        B_0 &= \hphantom{-}\frac{\Bar{S}(\vec{k}) (\gamma^2 + \omega_0^2) - \langle \omega^2 \rangle_{\text{cl}}^C}{(\alpha - \gamma)^2 + \omega_0^2},\\
        B_1 &= \hphantom{-}\frac{\Bar{S}(\vec{k}) \alpha (\alpha - 2\gamma) + \langle \omega^2 \rangle_{\text{cl}}^C}{(\alpha - \gamma)^2 + \omega_0^2},\\
        B_2 &= - \frac{\Bar{S}(\vec{k}) \alpha (\gamma^2 - \alpha\gamma - \omega_0^2) + \langle \omega^2 \rangle_{\text{cl}}^C (\alpha - \gamma)}{\omega_0 \left[ (\alpha - \gamma)^2 + \omega_0^2 \right]},
    \end{align}%
    \label{eq:GCM_constraints}%
\end{subequations}%
where $\Bar{S}(\vec{k}) = S(\vec{k}) - C$ and $\langle \omega^2 \rangle_{\text{cl}}^{C} = (1 - C)\vec{k}^2 \boltz T / m$, which is the extensions of the formulation in Ref.~\cite{wax2013effective} with the addition of a single-particle mode. This fitting procedure was based on the Fourier transform of the classical computation in equation \eqref{eq:classical_ISF}, formulated in terms of mass centres of the wave packets, as a description based on the full density from the wave packets include a weak high-frequency mode originating from the confinement potential. 

The semiclassical quantum correction is implemented by first shifting the resonance according to equation \eqref{eq:collective_recoil} and then the two contributions $S^{\text{GCM}}$ and $S^{\text{sp}}$ are corrected according to equation \eqref{eq:DB_correction_classical} and \eqref{eq:DB_correction_single_particel} respectively. This method automatically satisfies the $f$-sum rule: this is obvious for the single-particle term and for the collective term this is enforced by equations \eqref{eq:GCM_constraints} and \eqref{eq:DB_correction_classical}.

%%%%%%%%%%%%%%%%%%%%% Response coefficients $f$-sum %%%%%%%%%%%%%%%%%%%%%
\section{$f$-sum rule for density response functions}\label{app:f_sum_chi}

The imaginary part of the density response function $\Im\left\{ \chi_{\alpha\alpha}(\vec{k}, -\vec{k}, \omega) \right\}$ has an equivalent $f$-sum rule to the dynamic structure factor $S_{\alpha\alpha}(\vec{k}, \omega)$. Using the fluctuation-dissipation theorem \eqref{eq:fluctuation-dissepation} and $\Im\left\{ \chi_{\alpha\alpha}(\vec{k}, -\vec{k}, \omega) \right\} = - \Im\left\{ \chi_{\alpha\alpha}(\vec{k}, -\vec{k}, -\omega) \right\}$ the $f$-sum rule is 
\begin{equation}
    \int_{-\infty}^{\infty}\!d\omega\; \omega \Im{\chi_{\alpha\alpha}(\vec{k}, -\vec{k}, \omega)} = -\frac{2\pi N}{\hbar} \omega_R,
\end{equation}
where the $\hbar$ dependence cancel. To make a connection to the time domain one should account for the causal nature of the response coefficient, 
\begin{equation}
    \chi_{\alpha\alpha}(\vec{k}, -\vec{k}, t) = \Theta(t) \Bar{\chi}_{\alpha\alpha}(\vec{k}, -\vec{k}, t),
\end{equation}
where $\Theta(t)$ is the step function and $\Bar{\chi}_{\alpha\alpha}(\vec{k}, -\vec{k}, t)$ is an odd function in $t$. The relation to the frequency domain is now 
\begin{equation}
    \begin{aligned}
        \Bar{\chi}_{\alpha\alpha}(\vec{k}, &-\vec{k}, t)\\
        &= \frac{i}{\pi} \int_{-\infty}^{\infty}\!\!d\omega\; e^{-i\omega t} \Im{\chi_{\alpha\alpha}(\vec{k}, -\vec{k}, \omega)},
    \end{aligned}
\end{equation}
and the $f$-sum rule is
\begin{equation}
    \frac{d}{dt} \Bar{\chi}_{\alpha\alpha}(\vec{k}, -\vec{k}, t)|_{t=0} = -N\frac{\vec{k}^2}{m_{\alpha}},
\end{equation}
where $m_{\alpha}$ is the mass of the particles in question.

Looking specifically at the impulse response computations for the wave-packet formulation, where the particles have some classical position $\vec{r}_i$ and momentum $\vec{p}_i$ as well as some internal degrees of freedom $\vec{\sigma}_i$ with associated momentum variables $\vec{\pi}_i$. The single-particle density $|\phi_i(\vec{x})|^2$ is described by some envelope $\rho_i$, i.e.
\begin{equation}
    |\phi_i(\vec{x})|^2 = \rho_{i}(\vec{x} - \vec{r}_i; \vec{\sigma}_i).
\end{equation}
Introducing the external impulse $V^{\text{ext}}_{\alpha}(\vec{x}) = I_0 \delta(t) \exp(i\vec{k}\cdot\vec{x}) \delta_{\alpha e}$ the associated change in momentum variables are
\begin{subequations}
    \begin{align}
        \Delta \vec{p}_i &= -i I_0 e^{i\vec{k}\cdot\vec{r}_i} \theta_i(\vec{k}) \vec{k},\\
        \Delta \vec{\pi}_i &= - I_0 e^{i\vec{k}\cdot\vec{r}_i} \frac{\partial \theta_i(\vec{k})}{\partial \vec{\sigma}_i},
    \end{align}
\end{subequations}
and 
\begin{equation}
    \theta_i(\vec{k}; \vec{\sigma}_i) = \int d^3\vec{x}\; \rho_i(\vec{x}; \vec{\sigma}_i)\, e^{i\vec{k}\cdot\vec{x}}.
\end{equation}
For a short time after the impulse at $t=0$, the generalised trajectories are described by 
\begin{subequations}
    \begin{align}
        \vec{r}_i^{I_0}(t) &= \vec{r}_i(t) + \frac{\Delta \vec{p}_i}{m_{\alpha}} t + \order{t^2}\\
        \vec{\sigma}_i^{I_0}(t) &= \vec{\sigma}_i(t) + \frac{\Delta \vec{\pi}_i}{m_{\alpha}} t + \order{t^2} 
    \end{align}
\end{subequations}
where $\vec{r}_i(t)$ and $\vec{\sigma}_i(t)$ are the unperturbed trajectories. The density response may now be explicitly evaluated to $\order{t}$, 
\begin{equation}
    \begin{aligned}
        \Bar{\chi}_{\alpha\alpha}(\vec{k}, -\vec{k}, t) &= \frac{\langle \delta n(\vec{k}, t) \rangle}{I_0}\\
        &= \frac{1}{I_0}\left\langle \sum_{i} \theta_i(-\vec{k}; \vec{\sigma}_i^{I_0}) e^{-i\vec{k}\cdot\vec{r}_i^{I_0}} \right\rangle,
    \end{aligned}
\end{equation}
and, in particular, assuming an uncorrelated unperturbed motion
\begin{equation}
    \begin{aligned}
        \frac{d}{dt} \Bar{\chi}_{\alpha\alpha}(\vec{k}, -\vec{k}, t)|_{t=0} = -\frac{N}{m_{\alpha}} \Big( \vec{k}^2 &\left\langle \big|\theta_i(\vec{k})\big|^2 \right\rangle\\
        + &\left\langle\Big|\frac{\partial \theta_i(\vec{k})}{\partial \vec{\sigma}_i}\Big|^2 \right\rangle\Big)
    \end{aligned}
\end{equation}
which agrees with the $f$-sum rule for point particles $\theta_i(\vec{k}) = 1$ but the distributed nature of the particles effectively reduces the strength of the impulse when the wavelength of the impulse is comparable to the size of the wave packet. 

%%%%%%%%%%%%%%%%%%%%% Response coefficients f-sum %%%%%%%%%%%%%%%%%%%%%
\section{Impulse simulations and linearity of response}\label{app:impulse_linearity}
\begin{figure*}
    %\subfloat{\includegraphics[width=0.5\linewidth]{Figures/impulse_response_linear_long.jpg}}
    %\subfloat{\includegraphics[width=0.5\linewidth]{Figures/impulse_response_linear_short.jpg}}
    \includegraphics[width=\linewidth]{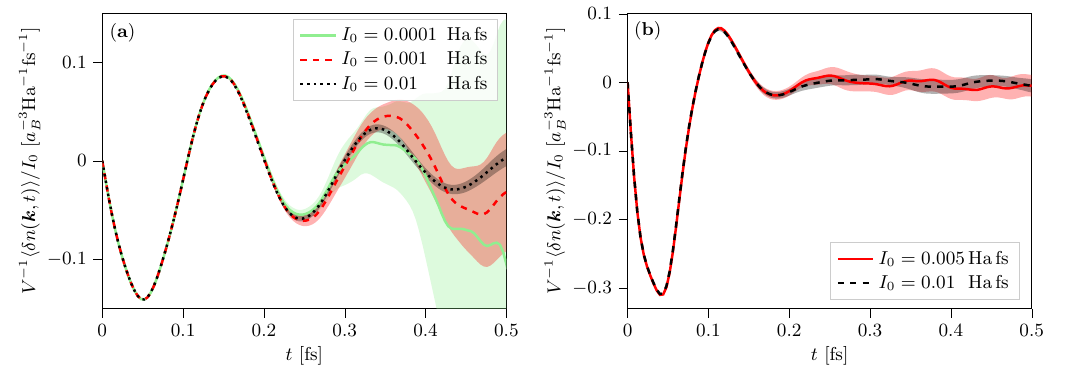}
    \caption{The linearity of the impulse response is tested for $(\lambda_s \vec{k})^{-1} = 2.1$ (a) and $(\lambda_s \vec{k})^{-1} = 1.1$ (b), by varying the impulse strength $I_0$. Shaded areas represent an estimate of the 95\% confidence internal assuming all the simulation results are Gaussian distributed around the response function. All simulations are seen to agree until the noise becomes dominant in which all simulations agree within the error estimates.}
    \label{fig:linear_impulse}
\end{figure*}%
\begin{table}
    \centering
    \caption{\label{tab:impulse}Summary of all impulse simulations shown for the two wavelength considered. A variety of impulse strengths $I_0$ were tested for a subset of the $360$ starting configurations to estimate nonlinear effects. The required simulation time $t_{\text{max}}$ for the shorter wavelength is reduced.}
    \begin{ruledtabular}
        \begin{tabular}{cc|ccc}
            $\vec{k}$ [$\bohr^{-1}$] & $I_0$ [$E_h\, \text{fs}$] & Start config. & Simulations & $t_{\text{max}}$ [fs] \\\hline
            \multirow{3}{*}{$0.31$} & $10^{-4}$ & 30 & 90 & $2$\\
            & $10^{-3}$ & 30 & 90 & $2$\\
            & $10^{-2}$ & 120 & 360 & $4$\\\hline
            \multirow{2}{*}{$0.61$} & $5\times 10^{-3}$ & 60 & 180 & $4$\\
            & $10^{-2}$ & 120 & 360 & $2$
        \end{tabular}
    \end{ruledtabular}
\end{table}%
The density response calculations where an explicit impulse is introduced -- as discussed in section \ref{sec:impulse_computations} -- are started from thermal simulations taken from a standard data collection run where each starting configuration was taken $20\,\text{fs}$ apart, where $24$ stating configurations were considered. These considerations were repeated for five different seed configurations, resulting in $120$ initial configurations to which the impulses were computed. Furthermore, for the $\vec{k}$ vectors considered, three symmetry directions exist and the impulses were performed in each direction separately. High-resolution data were also acquired without any impulse, and for varying impulse strengths to confirm the simulation was performed in the linear regime. A summary of the simulations is presented in table \ref{tab:impulse} where a total of $1080$ impulse simulations were performed.

The linearity of the impulse response required for the validity of equation \eqref{eq:density_resposne} is demonstrated in Fig.~\ref{fig:linear_impulse}. The responses (normalised with the impulse strength $I_0$) agree for substantially different $I_0$ demonstrating that the impulses were performed in the linear regime. For longer times after the impulse the different perturbations to the system start to deviate from one another -- however, always within the estimated error based on the variations between different starting configurations -- as the weaker impulses converge slower (with respect to averaging) and less data were collected for these cases when linearity was established. For the shorter wavelength case nonlinear effects could be seen at an order of magnitude larger impulses than shown here. Figure~\ref{fig:comparison_real} in the main text use an impulse strength of $I_0 = 0.01\,\text{Ha}\,\text{fs}$.

%%%%%%%%%%%%%%%%%%%%% BIBLIOGRAPHY %%%%%%%%%%%%%%%%%%%%%
% \nocite{*}
\bibliography{ref} % Produces the bibliography via BibTeX.

\end{document}